\documentclass[aps,prxq,twocolumn,groupedaddress,floatfix,sort&compress,longnamesfirst]{revtex4-2}%revtex4-2
\usepackage{float}
\usepackage{graphicx}% Include figure files
\usepackage{amsmath}

\usepackage{color,amsthm,amsmath,amsfonts,graphicx,bm,epsfig,float,siunitx,multirow,xr,enumerate,epstopdf,soul}
\usepackage[breaklinks=true,colorlinks=true,linkcolor=blue,urlcolor=blue,citecolor=blue]{hyperref}
\usepackage[T1]{fontenc}

\begin{document}

\title{Ultrastrong photon superbunching from electron shelving and time integral}

\author{He-bin Zhang$^{1}$}
\author{Yuanjiang Tang$^{1}$}
\author{Yong-Chun Liu$^{1,2}$}
\email{ycliu@tsinghua.edu.cn}
\affiliation{$^{1}$State Key Laboratory of Low-Dimensional Quantum Physics, Department of Physics, Tsinghua University, Beijing 100084, China}
\affiliation{$^{2}$Frontier Science Center for Quantum Information, Beijing 100084, China}

\date{\today}
	
\begin{abstract}
Photon correlation is at the heart of quantum optics and has important applications in quantum technologies. Here we propose a universally applicable mechanism that can generate the superbunching light with ultrastrong second-order and higher-order correlations hitherto unreachable. This mechanism arises from the combined effect of electron shelving and time integral of fluorescence based on a cascaded quantum system comprising an emitter and a filter or a cavity QED system, and has high experimental feasibility according to current experimental techniques. Besides, both the correlation degrees and the frequency of the light can be flexibly varied over broad ranges. Both the research and technological applications on strong correlations can be extensively facilitated due to this readily accessible and manipulated mechanism for generating photon correlation.  
\end{abstract}

\maketitle

\section{Introduction}
	
Exploration on various correlations of photons is the key to deep insight into physical laws~\cite{Einstein1935_Can, Bell1964_On, Aspect1981_Experimental, Glauber1963_The_Quantum, Glauber1963_Photon}, and has led to a large number of advanced technologies, including quantum communication~\cite{Bennett1984_Quantum}, quantum computing~\cite{Knill2001_A-scheme}, quantum metrology~\cite{Giovannetti2006_Quantum}, and quantum imaging~\cite{Shih2020_An}. Since proposed by Glauber in his pioneering work~\cite{Glauber1963_The_Quantum, Glauber1963_Photon}, the correlation function, which gauges the degree of photon statistical correlations, has taken center stage in quantum optics. For example, the single-photon emission satisfies a vanishing probability of multiphoton emission, i.e., the normalized second-order correlation function $g^{(2)}(0)=0$, which is indispensable for quantum communication~\cite{Bennett1984_Quantum, Bennett1992_Experimental} and optical quantum computing~\cite{Knill2001_A-scheme, O'Brien2009_Photonic}. There naturally exists the light field with perfect single-photon character, i.e., resonance fluorescence of a single emitter. Moreover, this simplest quantum field can be emitted from a variety of physical systems, such as atom~\cite{Kimble1977_Photon, Grangier1986_Observation}, quantum dot~\cite{Michler2000_A-Quantum}, molecule~\cite{Lounis2000_Single}, and color center~\cite{Kurtsiefer2000_Stable}. Therefore, resonance fluorescence has been widely used in research, which greatly promotes the development of quantum optics and quantum information science.  
	
As another extreme case of photon correlation, the light with second-order correlation $g^{(2)}(0) \gg 2$, i.e., superbunching field~\cite{Ficek2005_Quantum}, has also received tremendous research attention~\cite{Leon2019_Photon, Manceau2019_Indefinite-Mean, Ye2022_Antibunching, Spasibko2017_Multiphoton, Schaeverbeke2019_Single-Photon, Marconi2018_Far-from-Equilibrium, Valle2013_Distilling, Munoz2014_Emitters, Bin2021_Parity-Symmetry-Protected, Liu2016_Enhanced, Jahnke2016_Giant, Meuret2015_Photon, Boitier2011_Photon, Zhang2019_Superbunching, Zhou2017_Superbunching, Super-bunching2024_Qin} for the significant research value. The schemes for generating superbunching field and $n$-photon bundles were proposed based on the distillation of subsequent photon emissions~\cite{Munoz2014_Emitters, Valle2013_Distilling, Bin2021_Parity-Symmetry-Protected} and extended to phonon~\cite{Bin2020_N-Phonon}. Recent experiments observed significant superbunching effects induced by a nonlinear process pumped by bright squeezed vacuum (BSV)~\cite{Manceau2019_Indefinite-Mean, Spasibko2017_Multiphoton}. Moreover, strong photon correlation has been used in the research of quantum imaging~\cite{Ye2022_Antibunching, Shih2020_An, Zhang2019_Superbunching}, nonlinear optics~\cite{Chang2014_Quantum}, nonlinear light-matter interaction~\cite{Jechow2013_Enhanced, Mouloudakis2020_Pairing, Agarwal1970_Field-Correlation}, microscopy~\cite{Lubin2022_Photon}, and detection of weak interaction~\cite{Carreno2015_Exciting, Bin2018_Detection}. Regrettably, light with strong photon correlation is challenging to generate, and as the correlation grows stronger, its acquisition becomes increasingly elusive, thereby significantly impeding research on strong photon correlations.
	
In this paper, we propose a universally applicable mechanism for generating the light with ultrastrong second-order and higher-order correlations hitherto unreachable under ordinary conditions. This mechanism utilizes a filter to collect the photons emitted from a $\Lambda$-shape quantum emitter. Through the electron shelving effect manipulated by external fields, the population in the target transition channel can be significantly smaller than the total population of the emitter. Consequently, the emission can be converted into light with ultrastrong correlation under the time integral by a filter with an appropriate bandwidth. 
A most remarkable feature of this mechanism is that all the normalized second-order and higher-order correlations can readily reach unprecedented degrees in broad parameter ranges, compared to the strongest superbunching reported to date~\cite{Manceau2019_Indefinite-Mean, Meuret2015_Photon, Spasibko2017_Multiphoton, Munoz2014_Emitters, Valle2013_Distilling, Super-bunching2024_Qin}. Furthermore, both the degrees of correlations and the frequency of the light can be easily varied over broad ranges by manipulating external fields or filter. This mechanism can be generalized to various systems, as demonstrated using an actual physical system, and has high experimental feasibility due to the maturity of filtering techniques in experiments~\cite{Akopian2006_Entangled, Hennessy2007_Quantum, Sallen2010_Subnanosecond, Ulhaq2012_Cascaded}. Additionally, we show that this mechanism can also be extended to cavity QED systems~\cite{Thompson1992_Observation, Boca2004_Observation, Aoki2006_Observation, Dayan2008_A-Photon, Park2006_Cavity, Yoshie2004_Vacuum, Hennessy2007_Quantum}.   
Given its ability to generate ultrastrong photon correlations of all orders under relaxed and easy-to-implement conditions, this mechanism can significantly facilitate the research on strong correlations in quantum optics and quantum technologies.
	
\section{Model and theoretical method}
	
\begin{figure}
    \includegraphics[draft=false, width=0.7\columnwidth]{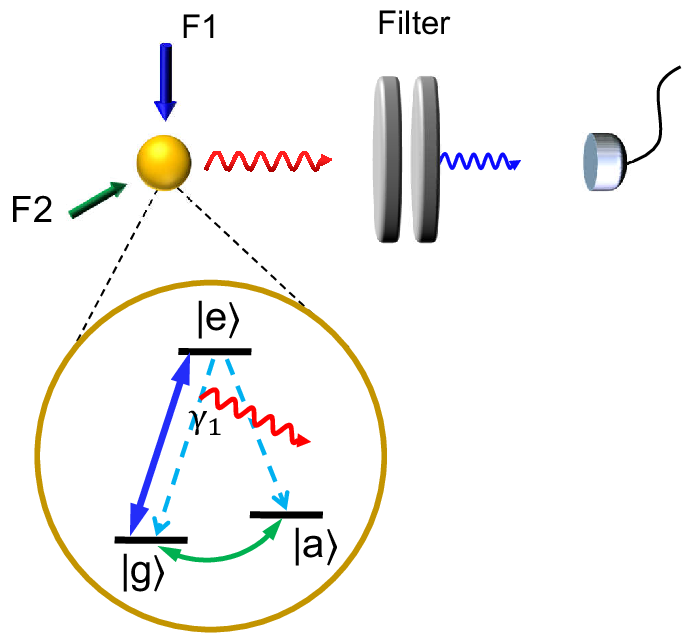}
	\caption{(Color online) Schematic of the cascaded quantum system consisting of a $\Lambda$-shape quantum emitter and filter-detector system. The emission from the transition $|e\rangle\rightarrow |g\rangle$ of the emitter is collected by the filter which is inserted between the emitter and the detector.
	\label{fig-1}}
\end{figure}
	
We consider a $\Lambda$-shape quantum emitter system, illustrated in Fig.~\ref{fig-1}, which includes an excited state $|e\rangle$ and two ground states $|g\rangle$ and $|a\rangle$. The transition $|g\rangle\leftrightarrow|e\rangle$ is driven by a laser field F1, and the transition $|g\rangle\leftrightarrow|a\rangle$ is driven by another coherent field F2. In the frame rotating at the driving field frequencies, the motion equation of this quantum emitter is given by
\begin{equation}
	\dot{\rho_e}  = -i[H_e,\rho_e] + \gamma_1 D[\sigma_{ge}]\rho_e + \gamma_2 D[\sigma_{ae}]\rho_e,
	\label{eq-1}
\end{equation}
where the Hamiltonian $H_e$ is defined as
\begin{equation}
	H_e = \Delta _a \sigma _{aa} + \Delta _e \sigma _{ee} + (\Omega \sigma_{eg} + \Omega_r \sigma_{ga} + \text{H.c.}).
	\label{eq-2}
\end{equation}
Here, $\sigma_{mn}$ represents the transition operator of emitter with $m,n=g,a,e$, and $\mathcal{D}[o]\rho\equiv o\rho o^\dagger-\frac{1}{2} o^\dagger o\rho-\frac{1}{2} \rho o^\dagger o$ is the Lindblad superoperator. 
${\Delta _e} = {\omega _e} - {\omega _{L1}}$ and $\Delta _a = {\omega _a} - {\omega _{L2}}$ denote the frequency detunings between the transitions $|g\rangle\leftrightarrow|e\rangle$ and $|g\rangle\leftrightarrow|a\rangle$, respectively, and the driving field F1 and F2. 
$\Omega$ and $\Omega_r$ denote the Rabi frequencies of the external fields F1 and F2, respectively. The decay rates from the excited state $|e\rangle$ to the ground states $|g\rangle$ and $|a\rangle$ are denoted by $\gamma_1$ and $\gamma_2$, respectively, with $\gamma_1=\gamma_2=\gamma$ for simplicity.

As depicted in Fig.~\ref{fig-1}, the emission from the transition $|e\rangle\rightarrow|g\rangle$ is collected by a filter with a specific center frequency and bandwidth. One of the quantities we focus on in this work is the zero-delay frequency-filtered correlation function~\cite{Knoll1990_Spectral, Nienhuis1993_Spectral} of this emission  
\begin{equation}
	g_b^{(N)}(0) = \frac{\langle \bar{\sigma}_{eg}(t)^{N}\bar{\sigma}_{ge}(t)^{N} \rangle}{\langle\bar{\sigma}_{eg}(t)\bar{\sigma}_{ge}(t)\rangle^{N}} .     \label{eq-3}
\end{equation}
The filtered field operator $\bar{\sigma}_{mn}$ is the convolution of the observable and the filtering function, and is denoted as 
\begin{equation}
	\bar{\sigma}_{mn}(t)=\int_{0}^{\infty}f(\tau)\sigma_{mn}(t-\tau)d\tau,  
	\label{eq-4}
\end{equation}
where $f(\tau)$ denotes filtering function. Due to causality, the integral over time $\tau$ is restricted to $\tau>0$. The form of the filtering function depends on the type of filter, with the Lorentzian filter being the most common:
\begin{equation}
	f(\tau)=e^{-(i\omega_{b}\tau+\frac{\kappa}{2})} ,
	\label{eq-5}
\end{equation}
where $\omega_{b}$ and $\kappa$ represent the filter's center frequency and bandwidth, respectively.

The computation of the frequency-filtered correlation function in Eq.~(\ref{eq-3}) involves multiple integrations, presenting significant computational challenges. However, the sensor method~\cite{Valle2012_Theory} offers an alternative approach to simplify these calculations.
Following this method, the filter, which is considered as a passive object that receives the emission from the quantum emitter without backaction on it, is modeled as a quantized harmonic oscillator with bosonic annihilation operator $b$.
Accordingly, the time evolution of this cascaded quantum system composed of the emitter and the filter as shown in Fig.~\ref{fig-1} is governed by the motion equation
\begin{equation}
	\dot{\rho}  = -i[H,\rho] + \gamma_1 D[\sigma_{ge}]\rho + \gamma_2 D[\sigma_{ae}]\rho + \kappa D[b]\rho.
	\label{eq-6}
\end{equation}
Here $\rho$ denotes the density operator of this cascaded quantum system, and the Hamiltonian $H$ is given by
\begin{equation}
	H = \Delta _a \sigma _{aa} + \Delta _e \sigma _{ee} + \Delta _b b^\dag b + \Omega \sigma_{eg} + \Omega_r \sigma_{ga} + g_c \sigma_{eg} b + \text{H.c.},
	\label{eq-7}
\end{equation}
where $\Delta _b = \omega _b - \omega _{L1}$ represents the detuning between the filtering frequency $\omega_b$ and the frequency $\omega_{L1}$ of field F1. For simplicity and without loss of generality, we next consider ${\Delta _a} = {\Delta _e} = {\Delta _b} = 0$, which indicates that both the transitions $|g\rangle\leftrightarrow|e\rangle$ and $|g\rangle\leftrightarrow|a\rangle$ are resonantly driven by external fields, and the filter is resonantly coupled to the emission from the transition $|e\rangle\rightarrow|g\rangle$ with coupling coefficient $g_c$. $g_c \rightarrow0$ ensures that the backaction of the filter on the emitter is negligible. The filtering bandwidth $\kappa$ determines the frequency resolution of the filter, and its inverse $\kappa^{-1}$ reflects the time resolution. Thus, this model intrinsically accounts for the uncertainty principle when describing filtering. 
Accordingly, the frequency-filtered correlation function in Eq.~(\ref{eq-3}) can be equivalently represented in a simple form  
\begin{equation}
	g_b^{(N)}(0)=\lim_{g_c\to0}\frac{\langle b^\dagger(0)^N b(0)^N\rangle}{\langle b^\dagger(0)b(0)\rangle^N} ,
	\label{eq-8}
\end{equation}
in the steady-state limit of the emitter system.

\section{Photon superbunching}
	
\begin{figure}
 	\includegraphics[draft=false, width=0.95\columnwidth]{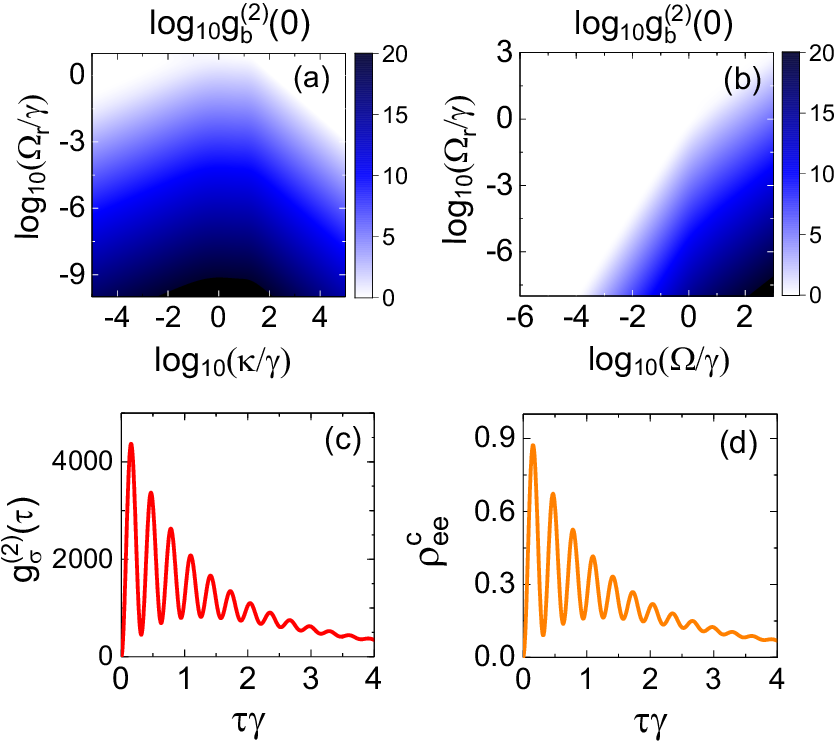}
	\caption{(Color online) Normalized frequency-filtered second-order correlation $ g_b^{(2)}(0)$ of the emission from the transition $|e\rangle\rightarrow|g\rangle$ (a) as functions of $\kappa$ and $\Omega_r$, with $\Omega=10\gamma$, and (b) as functions of $\Omega$ and $\Omega_r$, with $\kappa =\gamma$.
	(c) Normalized second-order correlation $g^{(2)}_{\sigma}(\tau)$ of the emission from the transition $|e\rangle\rightarrow|g\rangle$ and (d) conditional excitation $\rho_{ee}^c(\tau)$ of the emitter as functions of delay $\tau$, with $\Omega =10\gamma$ and $\Omega_r =10^{-1}\gamma$. 
	\label{fig-2}}
\end{figure}

In Figs.~\ref{fig-2}(a) and (b), we present the frequency-filtered correlation $g_b^{(2)}(0)$ as functions of $\Omega $, $\Omega_r$, and $\kappa $. As illustrated in the upper right corner of Fig.~\ref{fig-2}(a), when $\kappa\gg\gamma$, $g_b^{(2)}(0)$ approaches zero, which indicates the single-photon character of the fluorescence. As $\kappa$ decreases, the single-photon character gradually diminishes. When $\kappa\ll\gamma$, the filtered field is mainly composed of the incoherent components of fluorescence, and thus exhibits a Poissonian statistics. 
Notably, when $\Omega\gg\Omega_r$, the correlation $g_b^{(2)}(0)$ displays an extremely large value, indicating that the emission from the transition $|e\rangle\rightarrow|g\rangle $ is converted into a superbunching field after filtering. The degree of superbunching can be easily varied by manipulating the filtering bandwidth or the intensities of the external fields. Furthermore, the correlation $g_b^{(2)}(0)$ can achieve an unprecedented degree compared to previously reported studies~\cite{Manceau2019_Indefinite-Mean, Meuret2015_Photon, Spasibko2017_Multiphoton, Munoz2014_Emitters, Valle2013_Distilling, Super-bunching2024_Qin}.

\subsection{Origin of superbunching effect} 
	
To comprehend the origin of the superbunching effect in the filtered field, it is essential to investigate the correlation property of the emission. As shown in Fig.~\ref{fig-2}(c), the normalized second-order correlation $g^{(2)}_{\sigma}(\tau)$ of the emission from the transition $|e\rangle\rightarrow|g\rangle $ is zero when $\tau=0$, which indicates the perfect single-photon character of the fluorescence. Subsequently, $g^{(2)}_{\sigma}(\tau)$ rapidly increases to a remarkably large value as $\tau $ increases. To understand this phenomenon, we transform the expression of the unnormalized second-order correlation as $G_\sigma ^{(2)}(\tau ) = \mathop {\lim }\limits_{t \to \infty } \left\langle {{\sigma _{eg}}(t){\sigma _{eg}}(t + \tau ){\sigma _{ge}}(t + \tau ){\sigma _{ge}}(t)} \right\rangle  = {\tilde \rho _{ee}}\rho _{ee}^c(\tau ) $~\cite{Steck2024_Quantum}.
Here $\tilde{\rho}_{ee}$ is the steady-state excitation of the emitter, whose explicit expression is provided in Appendix~\ref{App-1}, and $\rho _{ee}^c(\tau ) \equiv {\rho _{ee}}(\tau ){|_{\rho (0) = \left| g \right\rangle \left\langle g \right|}}$ is the conditional excitation of state $|e\rangle$, which denotes the probability of the emitter evolving to the excited state $|e\rangle$ after a delay $\tau$ with the conditional (collapsed) state $|g\rangle$ as the initial state~\cite{Zhang2020_Monitoring}. 
Consequently, the normalized second-order correlation function can be reduced to
\begin{equation}
	g^{(2)}_{\sigma}(\tau)=\frac{\rho_{ee}^c(\tau)}{\tilde{\rho}_{ee} }.
	\label{eq-9}
\end{equation}
	
When $\Omega\gg\Omega_r$, state $|a\rangle$ becomes a shelving state~\cite{Nagourney1986_Shelved, Sauter1986_Observation, Bergquist1986_Observation, Plenio1998_The-quantum-jump} where the emitter's population predominantly resides, resulting in a significantly lower population in state $|g\rangle$ compared to the total population (See Appendix~\ref{App-1}). Due to the electron shelving in state $|a\rangle$, the steady-state excitation $\tilde{\rho}_{ee}$, i.e., the denominator in Eq.~(\ref{eq-9}), is small. Meanwhile, after a photon is emitted by the transition $|e\rangle\rightarrow|g\rangle$, the emitter will be determinately in the conditional state $|g\rangle$. Subsequently, the conditional excitation $\rho_{ee}^c(\tau)$ will evolve to a value significantly greater than $\tilde{\rho}_{ee}$ because the population in the conditional state $|g\rangle$ can be driven directly to the excited state $|e\rangle$ by the laser F1.
To illustrate this process, Fig.~\ref{fig-2}(d) shows the evolution of the conditional excitation $\rho_{ee}^c(\tau)$ with the initial state $|g\rangle$. It reveals that within a delay much shorter than the emitter's lifetime, the conditional excitation $\rho_{ee}^c(\tau)$ reaches its maximum value near 1, far exceeding the steady-state excitation $\tilde{\rho}_{ee}$. According to Eq.~(\ref{eq-9}), the normalized second-order correlation $g^{(2)}_{\sigma}(\tau)$ is determined by the ratio of $\rho_{ee}^c$ to $\tilde{\rho}_{ee}$, implying that the evolution patterns of $g^{(2)}_{\sigma}(\tau)$ and $\rho_{ee}^c$  with $\tau$ are similar, except for a proportionality factor $\tilde{\rho}_{ee}$ independent of $\tau$. Therefore, the evolution of $g^{(2)}_{\sigma}(\tau)$ in Fig.~\ref{fig-2}(c) can be understood by examining the evolution of $\rho_{ee}^c$ in Fig.~\ref{fig-2}(d).
The frequency-filtered second-order correlation $g^{(2)}_{b}(0)$ represents a convolution of the photons successively emitted from the transition $|e\rangle\rightarrow|g\rangle$ within the response time $\kappa^{-1}$ of the filter. Therefore, based on the evolution of $g^{(2)}_{\sigma}(\tau)$, we can infer that with an appropriate filtering bandwidth $\kappa$, $g^{(2)}_{b}(0)$ can exhibit a remarkably large value, indicating a significant superbunching effect.  
	
The aforementioned statistical properties of the emission and hence the filtered field stem from the condition that the population in the channel excited by the laser field under steady-state condition is much smaller than the total population of the emitter, i.e., $\tilde{\rho}_{gg}\ll 1$, which is satisfied over a broad parameter range (See Appendix~\ref{App-1}). Therefore, the superbunching effect arises under a relaxed condition, and the degree of superbunching effect can be easily adjusted by varying the filtering bandwidth or the intensities of the external fields F1 and F2, as demonstrated in Figs.~\ref{fig-2}(a) and (b). Additionally, we reveal that while maintaining the ultrastrong superbunching correlation, the frequency of superbunching field can be easily tuned over a broad range by altering the filtering frequency (See Appendix~\ref{App-3}).

\subsection{Effect of filtering bandwidth on superbunching}
\label{sec_sub2_2}
	
\begin{figure}[htbp]
	\includegraphics[draft=false, width=0.9\columnwidth]{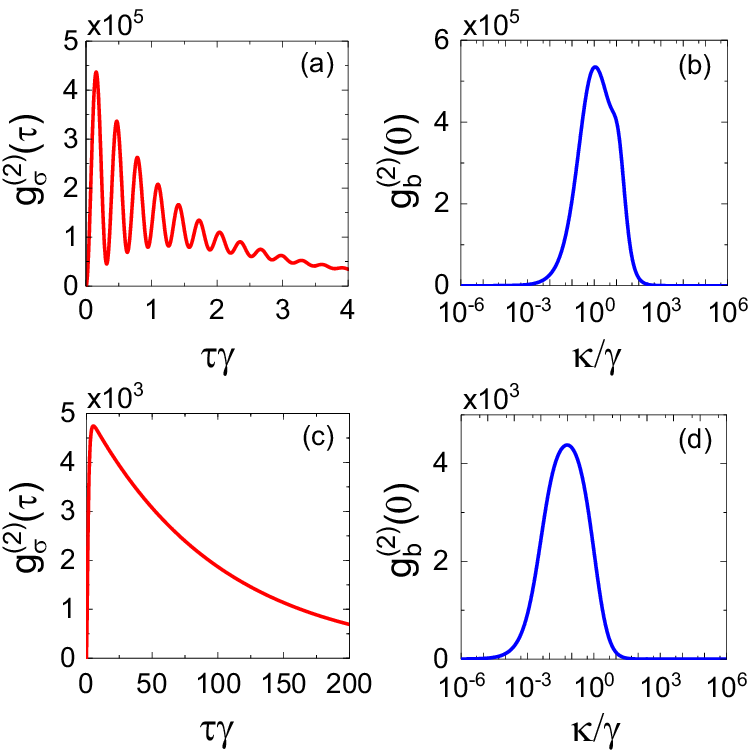}
	\caption{(Color online) Normalized second-order correlation $g^{(2)}_{\sigma}(\tau)$ as functions of delay $\tau$, with $\Omega=10\gamma, \Omega_r=10^{-2}\gamma $ in (a) and $\Omega=10^{-1}\gamma, \Omega_r=10^{-4}\gamma $ in (c). Normalized frequency-filtered second-order correlation $g^{(2)}_{b}(0)$ as functions of $\kappa$ in (b) and (d), with the same parameters as in (a) and (c), respectively.
	\label{fig-3}}
\end{figure}
	
In this section, we examine the influence of the different filtering bandwidth on the frequency-filtered correlation. As discussed earlier, there exists a close relationship between the frequency-blind correlation $g^{(2)}_{\sigma}(\tau)$ and the frequency-filtered correlation $g_b^{(2)}(0)$. Therefore, we can analyze the influence of filtering bandwidth on $g_b^{(2)}(0)$ by examining the property of $g^{(2)}_{\sigma}(\tau)$.
	
Under the strong-excitation regime of the laser field F1, i.e., $\Omega\gg\gamma$, the evolution of $g^{(2)}_{\sigma}(\tau)$ is shown in Fig.~\ref{fig-3}(a). It exhibits a damped harmonic oscillation with oscillation rate $\Omega$ and damping rate $\gamma$. The amplitude of oscillation near its maximum value is maintained for a duration corresponding to the excited-state lifetime $\gamma^{-1}$. Accordingly, when the response time of the filter is approximately equal to $\gamma^{-1}$, i.e., $\kappa\approx\gamma$, the convolution between the correlation of the emission and the filtering function, i.e., the frequency-filtered correlation $g_b^{(2)}(0)$, reaches its maximum $ max_ \kappa [g^{(2)}_{b}(0)]$. One can see from Figs.~\ref{fig-3}(a) and (b) that this maximum is of the same order of magnitude as $max_\tau [g^{(2)}_{\sigma}(\tau)]$ and thus $\tilde{\rho}_{ee}^{-1}$. Thereofre, the product of the frequency-filtered correlation and the intensity of emission reaches its upper limit (See Appendix~\ref{App-2}).
	
Under the weak-excitation regime of the laser field F1, i.e., $\Omega\ll\gamma$, the evolution of $g^{(2)}_{\sigma}(\tau)$ is shown in Fig.~\ref{fig-3}(c). The explicit expression of $g_\sigma^{(2)}(\tau)$ can be obtained as 
\begin{equation} 
	g_\sigma^{(2)}(\tau)=\frac{\Omega^2}{\tilde{\rho}_{ee}(\gamma^2-2\Omega^2)}(-2e^{-\gamma t} + e^{-2\gamma t} + e^{-\frac{2\Omega^2}{\gamma}t}).
	\label{eq-10}
\end{equation}
According to Fig.~\ref{fig-3}(c) and Eq.~(\ref{eq-10}), we can see that in the initial rising stage, the evolution rate of $ g_\sigma^{(2)}(\tau)$ mainly depends on $\gamma$. What follows is a monotonous decrease with a rate mainly determined by two evolution rates, i.e., $\frac{\Omega^2}{\gamma}$ and $\gamma$. We deduce that for this evolution, the optimal response time of the filter is determined by a compromise between these two evolution rates, i.e., their geometric mean $\sqrt{\gamma\Omega^2/\gamma}=\Omega$. Therefore, when $\kappa\approx\Omega$, the frequency-filtered correlation $g_b^{(2)}(0)$ reaches its maximum $max_\kappa [g^{(2)}_{b}(0)]$, which also has the same order of magnitude as $ max_ \tau [g^{(2)}_{\sigma}(\tau)]$ and thus $\tilde{\rho}_{ee}^{-1}$, as shown in Figs.~\ref{fig-3}(c) and (d).

\subsection{Higher-order correlation}
	
\begin{figure}
	\includegraphics[draft=false, width=0.8\columnwidth]{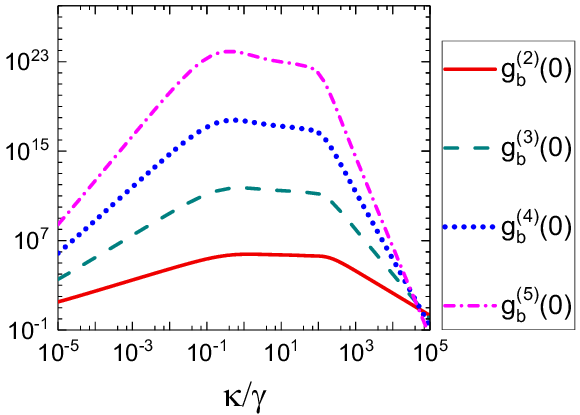}
	\caption{(Color online) Normalized frequency-filtered $N$-order correlation functions of the emission from the transition $|e\rangle\rightarrow |g\rangle$ as functions of $\kappa$ with $\Omega =10^{2}\gamma$ and $\Omega_r =10^{-1}\gamma$. 
	\label{fig-4}}
\end{figure}
	
In addition to the superbunching effect, the normalized frequency-filtered $N$-order correlation $g_b^{(N)}(0)$ of the emission from the transition $|e\rangle\rightarrow|g\rangle$ can also reach exceptionally large values as shown in Fig.~\ref{fig-4}. Furthermore, $g_b^{(N)}(0)$ increase exponentially with the order $N$. Similar to the case in second-order correlation, we find that $g_b^{(N)}(0)\approx \tilde{\rho}_{ee}^{1-N }$ when $\Omega \gg \gamma, \Omega_r$ and $\kappa \approx \gamma$. Therefore, the value of the $N$-order correlation $g_b^{(N)}(0)$ can be easily estimated according to the explicit expression of $\tilde{\rho}_{ee}$.
	
The origin of the exceedingly strong $N$-order correlations can be comprehended as follows. Following the emission of a target photon from the transition $|e\rangle\rightarrow|g\rangle$, the emitter in the ground state $|g\rangle$ can be rapidly driven back to the excited state $|e\rangle$ by the laser field F1, as demonstrated by the conditional evolution depicted in Fig.~\ref{fig-2}(d). Next, following another emission of the target photon, the emitter will be driven yet again by the laser field F1 to the excited state $|e\rangle$ and subsequently repeats the emission. Accordingly, a tight multi-photon sequence arises and is prominent in the photons emitted from the transition $|e\rangle\rightarrow|g\rangle$. This multi-photon sequence will be interrupted when the emitter reaches state $|a\rangle$ via either the dissipative channel $|e\rangle\rightarrow|a\rangle$ or the ground-state transition $|g\rangle\leftrightarrow|a\rangle$.  
Once the multi-photon sequence is interrupted, the emitter will enter a non-radiative period that is significantly longer than the period of the re-excitation driven by the laser F1, since the Rabi frequency of the field F2 driving the transition $|g\rangle\leftrightarrow|a\rangle$ is much weaker than that of the laser F1, i.e., $\Omega_r \ll \Omega$. Therefore, tight multi-photon sequences will constitute the main components of the emission from transition $|e\rangle\rightarrow|g\rangle$. 
When the multi-photon sequence passes through a filter with a finite time resolution of $\kappa^{-1}$, the temporal order of photons within the sequence cannot be distinguished by the filter, and thus the multi-photon sequence is converted into a temporally indistinguishable multi-photon cluster. Consequently, the emission is converted into a superbunching field with ultrastrong second-order and higher-order correlations after filtering.

\section{Possible implementations of the scheme}

\subsection{Implementation based on transition $F_{g}=1\rightarrow F_{e}=0$ of $D_{2}$ Line of ${}^{87}\rm{Rb}$.}
	
\begin{figure*}
	\includegraphics[draft=false, width=1.8\columnwidth]{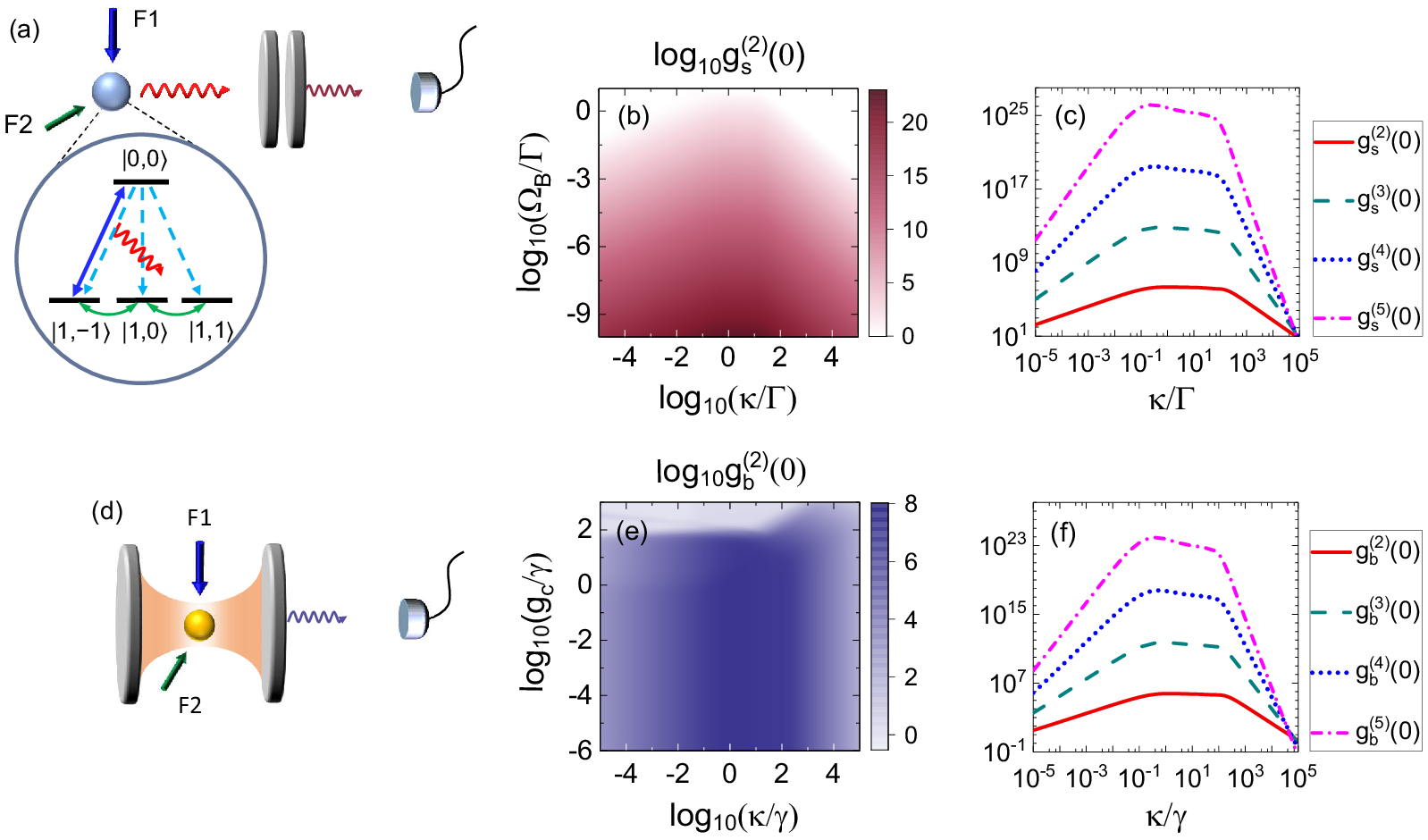}
	\caption{(Color online) (a) Scheme to generate superbunching field using $D_{2}$ transition of ${}^{87}$Rb atom as the emitter. The emission from the transition $|0,0\rangle\rightarrow|1,-1\rangle$ is collected by a filter.
	(b) Normalized second-order frequency-filtered correlation of this emission as functions of $\kappa$ and $\Omega_B$, with $V_{eg_{-1}} =10\Gamma$.  
	(c) Normalized $N$-order frequency-filtered correlation functions of this emission as functions of filtering bandwidth $\kappa$, with $V_{eg_{-1}} =10^{2}\Gamma$ and $\Omega_B =10^{-1}\Gamma$. Other parameters in (b) and (c) are  $\Delta_e=\Delta_s=0$.
	(d) Schematic diagram of the mechanism to generate superbunching field based on a cavity QED system. (e) Normalized second-order correlation of cavity field as functions of $\kappa$ and $g_c$, with $\Omega =10^{2}\gamma$ and $\Omega_r =10^{-2}\gamma$. (f) Normalized $N$-order correlations of cavity field as functions of $\kappa$ with $g_c =10^{-2}\gamma$, and other parameters are the same as in Fig.~\ref{fig-4}. 
	\label{fig-5}}
\end{figure*}
	
To demonstrate the universality and experimental feasibility of the proposed mechanism, we consider the transition $ F_{g}=1\rightarrow F_{e}=0$ of $D_{2}$ line of ${}^{87}$Rb as the emitter, as depicted in Fig.~\ref{fig-5}(a), where $F_g$ and $F_e$ represent the total angular momentums of the ground and excited states, respectively. This system could be implemented on various platforms, such as a single ${}^{87}$Rb atom loaded into an optical dipole trap, as reported in Refs.~\cite{Volz2006_Observation, Hofmann2012_Heralded, Leent2020_Long-Distance}. Similar to the configuration in Fig.1 in the main text, the transition $|1,-1\rangle \leftrightarrow |0,0\rangle$ is driven by a $\sigma^+$-polarized laser field, and the transitions between the ground-state Zeeman sublevels $ |1,0\rangle $ and $|1,\pm1\rangle$ are driven by a transverse magnetic field. The fluorescence emitted by the transition $|1,-1\rangle \leftrightarrow |0,0\rangle$ is collected by a filter.  
	
In the frame rotating at the laser frequency $\omega_L$ and the rotating wave approximation, the time evolution of the cascaded quantum system composed of the emitter and the filter is governed by the master equation
\begin{equation}
	\dot{\rho_r} =-\mathrm{i}[\textit{H}_{Rb},\rho_r]+\mathcal{L}_{\sigma}\rho_r  +\mathcal{L}_s \rho_r.
	\label{eq-11}
\end{equation}
Here the Hamiltonian of the combined system composed of the emitter and the filter is given by
\begin{equation}
	\textit{H}_{Rb}=\textit{H}_{0} + \textit{H}_{I} + \textit{H}_{B} + \textit{H}_{S}.     \label{eq-12}
\end{equation}
The nonperturbed Hamiltonian $\textit{H}_{0}$ of the emitter is given by
\begin{equation}
	\textit{H}_{0}=\Delta_e |F_{e},m_{e}\rangle \langle F_{e},m_{e}|+H.c .             \label{eq-13}
\end{equation}
where $\Delta_e$ is the detuning of the transition $|1,-1\rangle\leftrightarrow|0,0\rangle$ frequency from the laser.
The laser-atom interaction Hamiltonian $H_I$ is given by
\begin{equation}
	\textit{H}_{I}=\sum_{g_{i}} V_{eg_{i}} |F_{e},m_{e}\rangle \langle F_{g},m_{g_{i}}|+H.c .               \label{eq-14}
\end{equation}
According to the Wigner-Eckart theorem~\cite{Woodgate2000_Elementary, Brink1994_Angular, Meunier1987_A-simple}, the interaction energy for the transition $|F_{g},m_{g_i}\rangle\rightarrow|F_{e},m_{e}\rangle$ can be given by
\begin{widetext}
\begin{equation}
	V_{eg_{i}} = -\langle F_{e},m_{e}|\textbf{d}|F_{g},m_{g_{i}}\rangle\cdot\textbf{E} = (-1)^{F_{e}-m_{e}+1}\left(\begin{array}{ccc}F_{e} & 1  & F_{g} \\ -m_{e} & q & m_{g_{i}} \\
	\end{array}\right)\Omega_{L} ,              \label{eq-15}
\end{equation}
where $\textbf{d}$ denotes the electric dipole operator, $ \Omega_{L} = \left\langle F_{e}\parallel \textbf{d}\parallel F_{g}\right\rangle E $ represents the Rabi frequency of the light field, and $q=0, \pm1$ denotes the spherical components.
The magnetic-atom interaction Hamiltonian $H_{B}$ is represented as
\begin{equation}
	\textit{H}_{B} = \mu_{B} g_F\textbf{F}\cdot \textbf{B}_{T} = \sqrt{2}\Omega_{B} \sum_{g_{i},g_{j}} \langle F_{g},m_{g_{i}}|\textbf{F}|F_{g},m_{g_{j}}\rangle \cdot\textbf{e}_{x} |F_{g},m_{g_{i}}\rangle \langle F_{g},m_{g_{j}}| ,        \label{eq-16}
\end{equation}
with $\Omega_{B}=\mu_B g_F B_T/\sqrt{2}$, where $\mu_B$ and $g_F$ denote the Bohr magneton and the gyromagnetic factor of the ground states, respectively, and
\begin{equation}
	\langle F_{g},m_{g_{i}}|\textbf{F}|F_{g},m_{g_{j}}\rangle
	=\left\langle F\parallel \textbf{F}\parallel F\right\rangle (-1)^{F_{g}-m_{g_{i}}} \left(\begin{array}{ccc}F_{g} & 1  & F_{g} \\ -m_{g_{i}} & q & m_{g_{j}} \\
	\end{array}\right) .              \label{eq-17}
\end{equation}
\end{widetext}
The Hamiltonian $H_{S}$ describing the filter and its coupling with the emitter is given by
\begin{equation}
	\textit{H}_{S} = \Delta_{s} s^{\dag}s + (g_c \left|1,-1\right\rangle \left\langle 0,0\right|s^{\dag} + H.c ).                \label{eq-18}
\end{equation}
Here the first term represents the free Hamiltonian of the filter, with $\Delta_s=\omega_s-\omega_L$ denoting the frequency detuning between the filter and the laser. The second term represents the interaction between the emission and the filter. $\mathcal{L}_{\sigma}\rho_r$ represents the spontaneous decay of the emitter and takes the form
\begin{equation}
	\mathcal{L}_\sigma\rho_r=\sum_{i=-1}^{1}\Gamma_i D[\left|1,i\right\rangle\left\langle 0,0\right|]\rho_r.
	\label{eq-19}
\end{equation}
Here $\Gamma_i=\frac{\Gamma}{3}$ are the decay rates from excited state $|0,0\rangle$ to ground state $|1,i\rangle$ with $i=0,\pm 1$, and $\Gamma$ is the natural linewidth  of the transition $F_{g}=1\rightarrow F_{e}=0$ of the $D_{2}$ line of ${}^{87}$Rb. The dissipation term of the filter $\mathcal{L}_s\rho_r$ is given by
\begin{equation}
	\mathcal{L}_s\rho_r=\kappa\mathcal{D}[s]\rho_r .
	\label{eq-20}
\end{equation}
A more detailed mathematical description of the motion equation for this system can be found in Ref.~\cite{Zhang2019_Absorption}.
	
Analogous to the emitter shown in Fig.~\ref{fig-1}, when the driving fields and satisfies $V_{eg_{-1}}\gg\Omega_B$, states $|1,0\rangle$ and $|1,1\rangle$ are shelving states, and state $|1,-1\rangle$ is analogous to state $|g\rangle$ of the $\Lambda$-shape emitter.  Since the emitter's population is concentrated in the shelving states, the steady-state population in $|1,-1\rangle$  will be significantly smaller than the total population of the emitter. In Fig.~\ref{fig-5}(b), we plot the frequency-filtered second-order correlation of the emission from the transition $|1,-1\rangle\leftrightarrow|0,0\rangle$ as functions of $\kappa$ and $\Omega_B$. It reveals that the emission from the transition $|1,-1\rangle\leftrightarrow|0,0\rangle$ can also be converted into a superbunching field after filtering. 
	
Furthermore, we plot the frequency-filtered higher-order correlations of this emission in Fig.~\ref{fig-5}(c). It reveals that the normalized frequency-filtered higher-order correlations can also reach exceedingly large values, similar to the case shown in Fig.~\ref{fig-4}. Therefore, it can be inferred that using the emitter shown in Fig.~\ref{fig-5}(a), the superbunching effect with extremely strong second-order and higher-order correlations can be achieved, as the case presented in earlier sections.

\subsection{Cavity QED system}
	
To facilitate the study of frequency-filtered correlation, we adopt the formalism proposed in Ref.~\cite{Valle2012_Theory}, where the coupling coefficient $g_c$ between the emitter and the filter is mathematically required to approach zero. However, ignoring this requirement, the motion equation Eq.~(\ref{eq-6}) is identical to that of a cavity QED system as depicted in Fig.~\ref{fig-5}(d), which is a well-developed platform in experiments and can be implemented in various systems including Fabry-Perot cavity coupled to a single atom~\cite{Thompson1992_Observation, Boca2004_Observation}, whispering gallery mode coupled to a single atom~\cite{Aoki2006_Observation, Dayan2008_A-Photon} or NV color center in diamond~\cite{Park2006_Cavity}, and photonic-crystal cavity coupled to quantum dot~\cite{Yoshie2004_Vacuum, Hennessy2007_Quantum}. Given the wide applications of both strong photon correlation and cavity QED systems, it is worthwhile to investigate whether the aforementioned mechanism can be extended to cavity QED systems. 
Fig.~\ref{fig-5}(e) shows the second-order correlation of the cavity mode as functions of the cavity's bandwidth $\kappa$ and the coupling coefficient $g_c$. 
One can see that in the weak-coupling and intermediate-coupling regimes of the cavity QED system, i.e., $0 < {g_c} < \gamma $ and ${g_c} \sim \gamma $, the correlation function of the cavity field remains almost unchanged. This phenomenon indicates that the backaction of the cavity on the emitter, represented by the term  ${g_c}{\sigma _{eg}}b$ involved in Hamiltonian $H$ in Eq.~(\ref{eq-7}), has almost no influence on the generation of superbunching light, even though the strength of this backaction reaches a nonnegligible degree. The superbunching correlation weakens only when in the strong-coupling regime, i.e., ${g_c} \gg \gamma $. However, a significant superbunching effect can be maintained until the ultrastrong coupling regime is reached. Considering that achieving the strong-coupling condition of cavity QED systems requires advanced experimental techniques, the influence of the cavity's backaction to the emitter on the superbunching effect is negligible in typical experimental situations. A comparison between Fig.~\ref{fig-5}(f) and Fig.~\ref{fig-4} reveals that higher-order correlations can also maintain extremely large values, as in the emitter-filter system. Moreover, enormous second-order and higher-order correlations can arise when the bandwidth of the cavity mode lies within a wide range that is experimentally feasible.

\section{Discussion}  
	
Noteworthily, in our proposed mechanism, the effect played by the filter can be viewed as accumulating photons within the response time and converting them into multi-photon clusters. The parameter requirements for the physical systems, including the frequencies of the external fields, filtering frequency and bandwidth, are relaxed. Besides, even the filter with a very broad bandwidth can convert the emission into a field exhibiting a significant superbunching effect, as shown in Fig.~\ref{fig-2}(a). Although the superbunching effect can occur within a relaxed parameter regime, selecting specific parameters can optimize the quality of the superbunching field. For instance, under the strong-excitation regime of laser field F1, i.e., $\Omega \gg \gamma$, the frequency-filtered correlation reaches its maximum value when the filtering bandwidth satisfies $\kappa \approx \gamma$. Meanwhile, under the weak-excitation regime of laser field F1, i.e., $\Omega \ll \gamma$, the frequency-filtered correlation reaches its maximum value when $\kappa \approx \Omega$. By examining the relationship between the degree of superbunching and the intensity of the emission, we discover the optimal parameter condition, i.e., $\Omega \gg \gamma$ and $\kappa \approx \gamma$, where the product of the frequency-filtered correlation and the intensity of the emission reaches its upper limit (See Appendix~\ref{App-2}).

\section{Conclusion}
	
In summary, we propose a universally applicable mechanism for generating the light with ultrastrong second-order and higher-order correlations. We introduce this mechanism by considering a cascaded quantum system comprising a $\Lambda$-shape quantum emitter and a filter, and reveal that this ultrastrong correlation originates from the combined effect of electron shelving and time integral of fluorescence from the emitter . Furthermore, both the correlation degrees and the frequency of the light can be conveniently varied over broad ranges by manipulating external fields or filter. This mechanism can be extended to cavity QED systems and remains vaild for various emitter systems with different energy structures. 
Based on several main advantages, including ultrastrong correlation, easy availability of implementation conditions, and high adjustability, this mechanism will significantly advance the research on strong correlations in quantum optics and quantum technologies.

\begin{acknowledgments}
This work is supported by the National Key R\&D Program of China (Grant No. 2023YFA1407600), and the National Natural Science Foundation of China (NSFC) (Grants No. 12275145, No. 92050110, No. 91736106, No. 11674390, and No. 91836302).
\end{acknowledgments}

\appendix

\section{Steady-state solution of emitter}
\label{App-1}

In the limit of the vanishing coupling between the emitter and the filter, the filter acts as a passive object that receives the emission from the emitter without backaction. Therefore, we can independently consider and solve the motion equation~(\ref{eq-1}) describing the evolution of the emitter shown in Fig.~\ref{fig-1}. Under steady-state condition, we obtain the analytical solutions of the density matrix elements of the emitter as follows
\begin{align}
	\tilde{\rho}_{gg} &= \frac{\Omega_r^2(2\gamma^2 + \Omega^2 + 2\Omega_r^2)}{M} ,      \nonumber  \\
	\tilde{\rho}_{aa} &= \frac{\Omega^4 + \Omega_r^2(2\gamma^2 - \Omega^2 + 2\Omega_r^2)}{M} ,     \nonumber    \\
	\tilde{\rho}_{ee} &= \frac{2\Omega^2\Omega_r^2}{M} ,       \nonumber  \\
	\tilde{\rho}_{ge} &= \frac{2i\gamma\Omega\Omega_r^2}{M} ,     \nonumber    \\
	\tilde{\rho}_{ae} &= \frac{-\Omega^3\Omega_r + 2\Omega\Omega_r^3}{M} ,   \nonumber  \\
	\tilde{\rho}_{ga} &= \frac{-i\gamma\Omega^2\Omega_r}{M} ,
	\label{eq-A1}
\end{align}
where $M = \Omega^4  + 2 \Omega_r^2 (2\gamma^2 + \Omega^2 + 2\Omega_r^2)$ with $\gamma_1=\gamma_2=\gamma$.

\section{Emission intensity of superbunching  field }
\label{App-2}

\begin{figure*}[htbp]
	\includegraphics[draft=false, width=1.6\columnwidth]{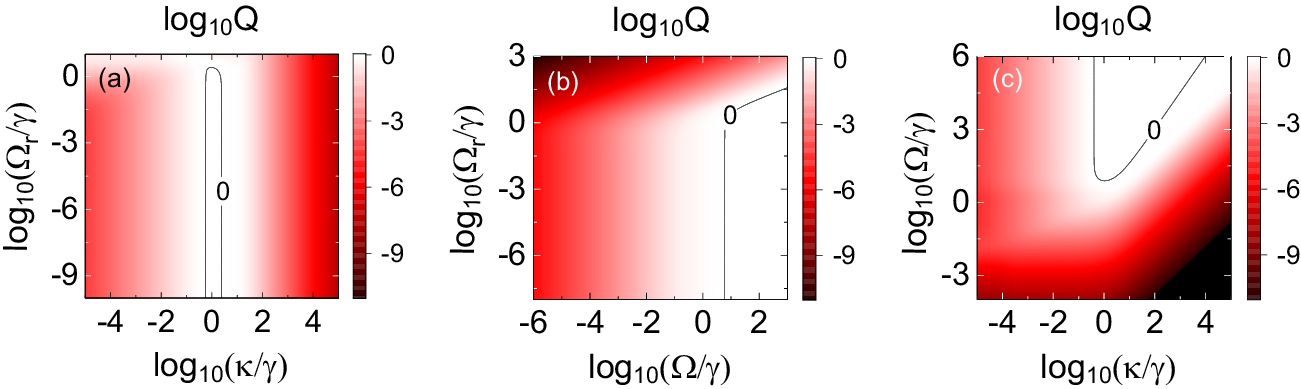}
	\caption{(Color online) Product $Q$ of the frequency-filtered correlation $g_b^{(2)}(0)$ and the steady-state excitation $\tilde{\rho}_{ee}$ of the emitter (a) as functions of $\kappa$ and $\Omega_r$ with $\Omega=10\gamma$; (b) as functions of $\Omega$ and $\Omega_r$ with $\kappa =\gamma$; (c) as functions of $\kappa$ and $\Omega$ with $\Omega_r =10^{-4}\gamma$. 
		\label{fig-6}}
\end{figure*}

To facilitate the understanding of the comprehensive property of the superbunching field, we consider the product of the frequency-filtered correlation and the excited population as a measurement reflecting the quality of the superbunching field, i.e., 
\begin{equation} 
	Q =  g_b^{(2)}(0) \tilde{\rho}_{ee}. 
	\label{{eq-A2}}
\end{equation} 
In Figs.~\ref{fig-6}(a)-(c), we show the quality $Q$ as functions of $\Omega$, $\Omega_r$, and $\kappa$. These figures reveal that in the parameter regime where the superbunching effect occurs (See Figs.~\ref{fig-2} (a)-(b)), $Q$ exhibits significant regularity. For instance, although the correlation $g_b^{(2)}(0)$ changes significantly when adjusting $\Omega_r$, $Q$ remains essentially constant. This phenomenon can be understood through the following analysis. According to Eq.~(\ref{eq-9}), we obtain
\begin{equation} 
	g_{\sigma}^{(2)}(\tau) \tilde{\rho}_{ee} = \rho_{ee}^c .
	\label{eq-A3}
\end{equation} 
The convolution of the term on the left side of the equal sign and the filtering function corresponds to $Q$. The term on the right side of the equal sign is the conditional excitation of emitter, which mainly reflects the conditional evolution of the subsystem composed of states $|g\rangle$ and $|e\rangle$ under the driving of the external field F1 and thus has nothing to do with the Rabi frequency $\Omega_r$ of the coherent field F2. Therefore, although adjusting $\Omega_r$ can significantly change the degree of superbunching (see Fig.~\ref{fig-2}(a) ), the quality $Q$ of the superbunching field remains unchanged. 

As discussed in Sec.~\ref{sec_sub2_2}, under the strong-excitation regime of the laser field F1, i.e., $\Omega \gg \gamma$, the correlation $g_b^{(2)}(0)$ reaches its maximum when $\kappa \approx \gamma$. Besides, when the external field F1 satisfies $\Omega \gg \gamma, \Omega_r$, the maximum amplitude of the conditional excitation $\rho_{ee}^c$ reaches its maximum value of 1, as shown in Fig.~\ref{fig-2}(d). Accordingly, when the conditions $\Omega \gg \gamma, \Omega_r$ and $\kappa \approx \gamma$ are simultaneously satisfied, the quality $Q$ of the superbunching field reaches its upper limit, i.e., $Q \approx 1$, as shown in the white regions in Figs.~\ref{fig-6}(a)-(c). 
In other words, although both the frequency-filtered correlation and the intensity of the emission (and thus the intensity of the filtered field) can be significantly changed by adjusting the intensities of the external fields F1 and F2, the frequency-filtered correlation always satisfies
\begin{equation} 
	g_b^{(2)}(0) \approx \tilde{\rho}_{ee}^{-1} ,
	\label{eq-A4}
\end{equation} 
when $\Omega \gg \gamma, \Omega_r$ and $\kappa \approx \gamma$.

Moreover, under the premise that $Q$ is close to its maximum value, we see from Fig.~\ref{fig-6}(c) that the adjustable range of $\kappa$ increases as $\Omega$ increases.
Therefore, we conclude that the three parameters $\Omega, \Omega_r, \kappa$ can be tuned over relaxed ranges while maintaining the quality $Q$ of superbunching field near its maximum, as shown in Figs.~\ref{fig-6}(a)-(c). 
Additionally, similar to second-order correlation, higher-order correlation also reaches the upper limit, i.e., 
\begin{equation} 
	g_b^{(N)}(0) \approx \tilde{\rho}_{ee}^{1-N},
	\label{eq-A5}
\end{equation} 
in the above parameter regime (See Fig.~\ref{fig-4}).

The emission intensity of a quantum emitter depends on the product of the maximum emission rate and the excited population. Although a certain correlation value limits the population excitation according to the above discussion, it does not limit the maximum emission rate of a quantum emitter. In fact, the maximum emission rate is determined by the species of the quantum system and varies enormously from species to species~\cite{Lounis2005_Single-photon}. For instance, the typical maximum emission rate of a quantum dot is about 5 orders of magnitude greater than that of a cold atom~\cite{Lounis2005_Single-photon}. Selecting a quantum emitter with a large emission rate, e.g., quantum dot, and setting the system to the optimal parameter region as described above, i.e., the white region satisfying $Q \approx 1$ in Fig.~\ref{fig-6}, it is possible to obtain a superbunching field with both a large field intensity and correlation degree. Consequently, based on an appropriate quantum system, one can obtain a superbunching field exceeding the strongest correlation achieved experimentally to date~\cite{Manceau2019_Indefinite-Mean, Spasibko2017_Multiphoton}, while maintaining the field intensity.

\section{Tunable frequency of superbunching field}
\label{App-3}

\begin{figure}[htbp]
	\includegraphics[draft=false, width=0.8\columnwidth]{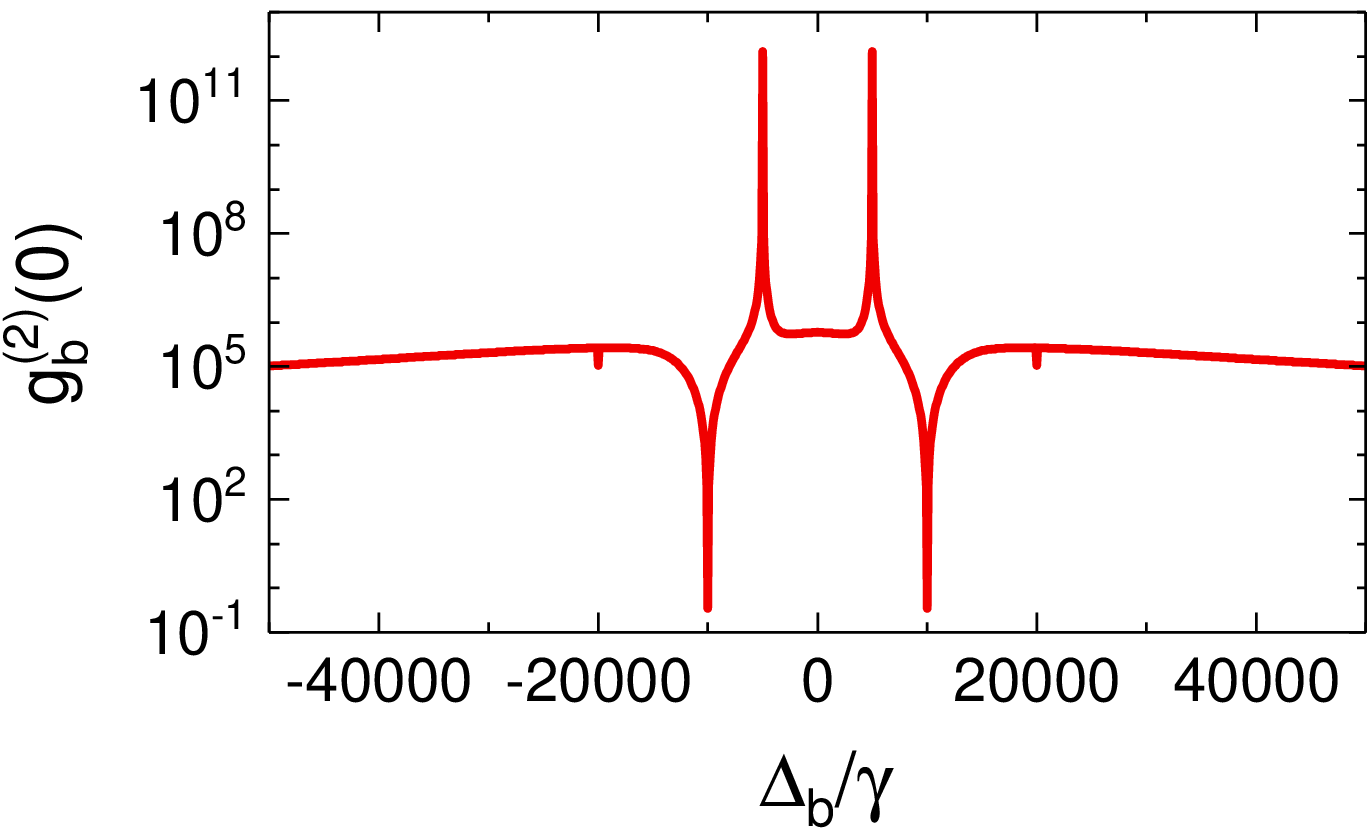}
	\caption{(Color online) Normalized frequency-filtered second-order correlation $ g_b^{(2)}(0)$ as a functions of $\Delta_b$ with $\Omega=10^{4}\gamma$, $\Omega_r =10\gamma$. 
	\label{fig-7}}
\end{figure}

The filtering frequency determines the center frequency of the filtered field. Previously, we mainly considered the case where the center frequency of the filtered field is equal to the frequency of the laser field F1, i.e., ${\Delta _b} = 0$. For the more general case of ${\Delta _b} \ne 0$, the total Hamiltonian of the system in Eq.~(\ref{eq-7}) becomes the following form
\begin{equation}
	H = \Delta_b b^\dagger b + (\Omega \sigma_{eg} + \Omega_r \sigma_{ga} + g_c \sigma_{eg} b + \text{H.c.}).
	\label{eq-A6}
\end{equation}

In Fig.~\ref{fig-7}, we show the frequency-filtered correlation  $ g_b^{(2)}(0)$ as a function of $\Delta_b$. It reveals that the ultrastrong superbunching correlation of the filtered field is well maintained when the filtering frequency is tuned over a broad range. 
Moreover, due to the effect of sidebands, changes in the correlation function arise at ${\Delta _b}= \pm \bar \Omega /2, \pm \bar \Omega, \pm 2\bar \Omega $, where $\bar\Omega=\sqrt {{\Omega ^2} + \Omega _r^2} $ denotes the dressed-state splitting. For instance, the coupling of the transition $\left| e \right\rangle  \leftrightarrow \left| g \right\rangle $ to the transition $\left| g \right\rangle  \leftrightarrow \left| a \right\rangle $ changes the correlation property of the emission from the transition $\left| e \right\rangle  \leftrightarrow \left| g \right\rangle $ in the sidebands at $\omega_{L1} \pm \bar \Omega$. The population entering into the shelving state $\left| a \right\rangle $ does not contribute to the target photon emission from the transition $\left| e \right\rangle  \leftrightarrow \left| g \right\rangle $, and even induces a blocking to this emission. Therefore, the mechanism for generating superbunching effect described earlier is disrupted at these sidebands, leading to the disappearance of the superbunching effect at $\Delta _b=\pm \bar\Omega $. However, the superbunching effect is well maintained at all other positions. 
Therefore, a superbunching field with ultrastrong correlation and tunable frequency is achieved.

\bibliographystyle{apsrev4-2}
\bibliography{bib_UltraStrong}

\end{document}